\documentclass[12pt]{iopart}

\usepackage{iopams}
\usepackage[utf8]{inputenc}
\usepackage[english]{babel}

\expandafter\let\csname equation*\endcsname\relax
\expandafter\let\csname endequation*\endcsname\relax 

\usepackage{amsmath}
\usepackage{amsfonts}
\usepackage{amssymb}
\usepackage{color}
\usepackage[dvipsnames]{xcolor}
\usepackage{mathrsfs}
\usepackage{graphicx}
\usepackage{lmodern}
\usepackage{lineno}
\usepackage{hyperref}
\usepackage{soul}
\usepackage{relsize}
\usepackage[all]{xy}

\newcommand{\seta}{\mathlarger{\eta}}
\newcommand{\kan}[2]{\overset{~\!\mathsmaller{#2}}{#1}}
\newcommand{\dubbra}{\mathrel{\{\! [ }}
\newcommand{\dubarb}{\mathrel{]\! \} }}

\newcommand{\gerst}[4]{\dubbra\kan{#1}{#3},\kan{#2}{#4}\dubarb}
\newcommand{\jacger}[3]{\dubbra\kan{#1}{#3},#2\dubarb}
\newcommand{\jacges}[3]{\dubbra #1,\kan{#2}{#3}\dubarb}
\newcommand{\codif}[2]{\mathrm{d}\bullet\kan{#1}{#2}}
\newcommand{\Hdw}{H_{\!\mathsmaller{\mathsmaller{DW}}}}
\newcommand{\intprod}{\mathrel{\!\mathlarger{\mathlarger{\mathlarger{\lrcorner}}} }}

\newcommand{\beq}{\begin{eqnarray}}
\newcommand{\eeq}{\end{eqnarray}}
\newcommand{\nn}{\nonumber}

\def\keywords#1{\vspace{10pt}
     \begin{indented}
     \item[]\rm Keywords: #1\par
     \end{indented}}

\begin{document}



\title{De Donder-Weyl Hamiltonian formalism of 
MacDowell-Mansouri gravity}

\author{Jasel Berra--Montiel$^{1}$, Alberto Molgado$^{1,2}$ and David Serrano-Blanco$^{1}$}

\address{$^{1}$ Facultad de Ciencias, Universidad Autonoma de San Luis 
Potosi \\
Av.~Salvador Nava S/N Zona Universitaria, San 
Luis Potosi, SLP, 78290 Mexico}
\address{$^{2}$ Dual CP Institute of High Energy Physics, Mexico}

\eads{\mailto{\textcolor{blue}{jberra@fc.uaslp.mx}},\ 
\mailto{\textcolor{blue}{molgado@fc.uaslp.mx}}, \ 
\mailto{\textcolor{blue}{davidserrano@fc.uaslp.mx}}
}


\begin{abstract}
We analyse the behaviour of the MacDowell-Mansouri action with internal symmetry group $\mathrm{SO}(4,1)$ under the De Donder-Weyl Hamiltonian formulation. The field equations, known in this formalism as the De Donder-Weyl equations, are obtained by means of the graded 
Poisson-Gerstenhaber bracket structure present within the De Donder-Weyl formulation. The decomposition of the internal algebra $\mathfrak{so}(4,1)\simeq\mathfrak{so}(3,1)\oplus\mathbb{R}^{3,1}$ allows the symmetry breaking $\mathrm{SO}(4,1)\to\mathrm{SO}(3,1)$, which reduces the original action to the Palatini action without the topological term. We demonstrate that, in contrast to the Lagrangian approach, this symmetry breaking can be performed indistinctly in the polysymplectic formalism either before or after the variation of the De Donder-Weyl Hamiltonian has been done, recovering Einstein's equations via the Poisson-Gerstenhaber bracket.
\end{abstract}




\section{Introduction}\label{intro}
The MacDowell-Mansouri model, first proposed in \cite{mm}, consists in a Yang-Mills-type gauge theory with gauge group $\mathrm{SO}(4,1)$. The relevance of this model relies in the fact that after the symmetry breaking $\mathrm{SO}(4,1)\rightarrow\mathrm{SO}(3,1)$ the action describing the gauge theory turns out to be classically equivalent to the standard Palatini action of General Relativity as both only differ by a topologically invariant term associated to the $\mathrm{SO}(3,1)$-curvature whose variation vanishes. The role of this model as a theory describing General Relativity has been vastly studied 
from different perspectives such as its BF reformulation \cite{wise,fri}, its connection with supergravity \cite{niet,niet2,cham}, its dual gravity formulation \cite{sab}, among others (see 
also~\cite{west,pegels,gotz}).

The equivalence between the action proposed by MacDowell and Mansouri and the Palatini action is made possible by the fact that the internal Lie algebra admits the orthogonal splitting $\mathfrak{so}(4,1)\simeq\mathfrak{so}(3,1)\oplus\mathbb{R}^{3,1}$, so that the symmetry breaking is achieved by projecting the associated $\mathrm{SO}(4,1)$-connection to its $\mathrm{SO}(3,1)$ components, which turn out to be the standard Lorentz connection. As pointed out by Wise \cite{wise}, a concise geometrical meaning can be given to the symmetry breaking process by identifying the $\mathrm{SO}(4,1)$ gauge field as a connection of a Cartan geometry \cite{cartan,cartan2}. 

Being a theory with a strong geometric background and also physically relevant due to its relation with General Relativity, the MacDowell-Mansouri model is a perfect candidate for analysis under the inherently geometric classical formulation known as the polysymplectic formalism.  
 Based on the early work of De Donder, Weyl, Carathéodory, among others \cite{donder,weyl,carath}, the polysymplectic approach of field theory consists in a De Donder-Weyl Hamiltonian formalism endowed with a generalization of the symplectic structure which enables to define a Poisson-like bracket. This issue, in particular, sets an important distinction with most 
of the various versions of the  multisymplectic formalism. 
 One of the key points within 
the polysymplectic approach relies in changing the definition of the standard Hamiltonian momenta, here considered not as  variations of the Lagrangian density with respect to time derivatives of the field variables but as variations with respect to the derivatives of the fields in every spacetime direction. These new fields, known within this context as polymomenta, define a De Donder-Weyl Legendre transformation which associates to the Lagrangian density a new function $\Hdw$ called De Donder-Weyl Hamiltonian, and it dictates the physical behaviour of the system in the so-called polymomentum phase-space. Without a space-like foliation, the polysymplectic formalism provides a manifestly spacetime covariant formulation. The polymomentum phase-space is endowed with a canonical $(n+1)$-form, called the polysymplectic form \cite{kan1,kan10,kan11,kan2}, which plays a similar role to
the one of the standard symplectic 2-form. A bracket may be induced in the polymometum phase-space by this polysymplectic form, which results to be a graded generalization of the Poisson bracket, called the Poisson-Gerstenhaber bracket \cite{kan2,kan6}. The Hamiltonian field equations 
encountered within 
this formalism, known as the De Donder-Weyl equations, can be written in terms of the
Poisson-Gerstenhaber bracket~\cite{kan1,kan10,kan11,kan2,kan6} with all spacetime variables treated on an equal
footing.

Our main purpose in this paper is to study the symmetry breaking process in the MacDowell-Mansouri action within the polysymplectic formalism. As we will describe below, we notice a significant difference between the Lagrangian and the De Donder-Weyl Hamiltonian approaches. In the former, the variation of the action and the symmetry breaking do not commute, thus the Lagrangian formalism is affected by the order in which we consider these two steps. In the latter, the field equations remain invariant irrespectively of the order in which these steps are considered. This 
invariance at the polysymplectic level results  naturally encoded by the definition of the polymomenta, as it contains information on the spacetime derivatives 
of all of the configuration variables, 
which remain unaltered whether or not the symmetry is broken.  
Indeed, the polymomenta can be defined as the variation of the action with respect to the spacetime derivatives of the configuration variables, thus, by breaking the $\mathrm{SO}(4,1)$ symmetry the intrinsic variation with respect to both variables present in the polymomenta remains after taking the projection down to $\mathfrak{so}(3,1)$, while in the Lagrangian case, when we break the symmetry, the variation with respect to 
some fields is lost in the process, thus, giving different equations.

At this point, we want to remark that our original goal 
is to apply the polysymplectic formalism to a physical motivated model. In
this sense, we thought that our main readership would be researchers with
a strong physical background and with a knowledge of mathematics at a
standard General Relativity level. In consequence, we deliberately keep the
more advanced mathematical issues of the formulation as simple as possible.

This paper is organized as follows: In 
Section~\ref{Mf} we review the main aspects of the polysymplectic approach in order to set up our notation and definitions, following as close as possible
references~\cite{kan1,kan10,kan11,kan2,kan6,kan3,kan4,kan5} in which a well behaved and intuitive framework  of the 
De Donder-Weyl Hamiltonian formalism is introduced. In section \ref{MM}, we introduce the MacDowell-Mansouri action following the notation developed in \cite{mm} and \cite{wise}. We describe the symmetry breaking $\mathrm{SO}(4,1)\rightarrow\mathrm{SO}(3,1)$ in detail for both, the Lagrangian and the polysymplectic formalisms.  In particular, we discuss on the different 
manner in which the symmetry breaking has to be interpreted 
within each formalism.  Finally, in Section~\ref{sec:conclu}
we include some concluding remarks.

\section{Polysymplectic formalism}\label{Mf}
In this section we briefly review the polysymplectic approach to the De Donder-Weyl formalism of classical field theory. We will closely follow the work of Kanatchikov \cite{kan1,kan10,kan11,kan2,kan6,kan3,kan4,kan5} on which the geometric and algebraic structures of the formalism have been vastly explored at the classical level.

Let us start by considering the fibre bundle $(E,\pi,M)$, with $M$ an $n$-dimensional oriented smooth manifold with local coordinates $\{x^{\mu}\}_{\mu=1}^{n}$, $E$ a smooth manifold of dimension $m$ with local coordinate functions $\{y^{a}\}_{a=1}^{m}$ and $\pi:E\rightarrow M$ the standard projection map. Let $\phi:M\rightarrow E$ be a local section around a point $p\in M$, so that its local components are given by $\phi^{a}:=y^{a}\circ\phi$. We will denote the space of local sections of $\pi$ around the point $p$ as $\Gamma_{p}(\pi)$. In what follows, 
we will consider the first jet manifold $J^{1}E$ of $\pi$, whose elements $j^{1}_{p}\phi\in J^{1}E$, for each point $p$, and every local section $\phi$, have the local representation $(x^{\mu},\phi^{a},\phi^{a}_{\mu})$, where $\phi^{a}_{\mu}:=\partial\phi^{a}/\partial x^{\mu}$ are commonly known as the derivative coordinates \cite{sardan,saund}. 

The behaviour of a physical system is described by the action functional $S:\Gamma_{p}(\pi)\rightarrow\mathbb{R}$, given in terms of the Lagrangian density $\mathcal{L}:J^{1}E\rightarrow\mathbb{R}$, such that $\mathcal{L}(j^{1}_{p}\phi)=\mathcal{L}(x^{\mu},\phi^{a},\phi_{\mu}^{a})$, by
\begin{equation}
\mathcal{S}[\phi]:=\int_{\mathcal{C}}\mathcal{L}(j^{1}_{p}\phi)\,V\,,
\end{equation} 
for any $\phi\in\Gamma_{p}(\pi)$, where $V:=dx^{1}\wedge\cdots\wedge dx^{n}$ denotes the local volume element of $M$ and $\mathcal{C}$ is an appropriate submanifold of $M$ on which the integration is performed. In order to determine the local sections which effectively describe the physical behaviour of a given theory, let us define the quantities
\begin{equation}\label{polym}
\pi_{a}^{\mu}:=\frac{\partial\mathcal{L}}{\partial\phi_{\mu}^{a}}\,,
\end{equation} 
called the \textit{polymomenta}. The triple $(x^{\mu},\phi^{a},\pi_{a}^{\mu})$ defines a local coordinate system for a new manifold $\mathcal{P}$, known in this context as the \textit{polymomentum phase-space}. 
In a heuristic way, the manifold $\mathcal{P}$ may be understood as a quotient bundle over $E$ 
associated to 
$(J^1E)^*$ (which stands for the affine dual bundle of 
the jet manifold $J^1E$) such that its tangent bundle structure allows a splitting into vertical and horizontal parts as described later on (see~\cite{kan6} for further details). 
The Lagrangian density $\mathcal{L}$ is associated to a smooth function $\Hdw$ in $\mathcal{P}$ by the map $\mathbb{L}:C^{\infty}(J^{1}E)\longrightarrow C^{\infty}(\mathcal{P})$, called the \textit{De Donder-Weyl Legendre transformation}, such that in local coordinates 
is explicitly given by
\begin{equation}\label{covleg}
\Hdw (x^{\mu},\phi^{a},\pi_{a}^{\mu}):=\mathbb{L}(\mathcal{L})=\pi_{a}^{\mu}\,\partial_{\mu}\phi^{a}-\mathcal{L}(x^{\mu},\phi^{a},\pi_{a}^{\mu})\,,
\end{equation}
where the symbol $\partial_{\mu}$ stands for the derivative with respect to the base space coordinates $x^{\mu}$. The function $\Hdw :\mathcal{P}\longrightarrow\mathbb{R}$ is called the \textit{De Donder-Weyl Hamiltonian} \cite{got,sardan2}.

Before we proceed, it is convenient to introduce the main geometrical objects associated to the polymomentum phase-space. Following standard notation, let $\mathfrak{X}(\mathcal{P})$ be the space of all sections of the tangent bundle $T\mathcal{P}$, that is, the space of smooth vector fields. Let $T^{V}\mathcal{P}$  denote the \textit{vertical tangent bundle} of $\mathcal{P}$, defined as the kernel of the pushforward of the bundle projection $\mathcal{P}\rightarrow M$, i.~e., 
the set of vectors which are not tangent to the 
base space manifold~\cite{saund}. The space of sections of the vertical tangent bundle, $\mathfrak{X}^{V}(\mathcal{P})$, will be regarded as the space of \textit{vertical vector fields}, which in local coordinates may be written as $X=X^{v}\partial_{v}:=X^{a}\partial_{a}+X^{\mu}_{a}\partial_{\mu}^{a}$, where we have adopted the notation $\partial_{a}:=\frac{\partial}{\partial\phi^{a}}$ and $\partial_{\mu}^{a}:=\frac{\partial}{\partial\pi_{a}^{\mu}}$ and the horizontal term is absent due to the 
definition above.

 Now, let $\Omega^{r}(\mathcal{P}):=\bigwedge^{r}\Omega^{1}(\mathcal{P})$, be the space of $r$-forms, 
that is, sections of $\bigwedge^{r}T^{*}\mathcal{P}$. The subspace $\Omega^{r}_{0}(\mathcal{P})\subset\Omega^{r}(\mathcal{P})$, whose elements satisfy $X\intprod\theta=0$ for $\theta\in\Omega^{r}_{0}(\mathcal{P})$ and for any $X\in\mathfrak{X}^{V}(\mathcal{P})$, is called the space of \textit{horizontal $r$-forms}. Finally, let us consider the space $\Omega_{1}^{r}(\mathcal{P})\subset\Omega^{r}(\mathcal{P})$ whose elements satisfy $X\intprod\theta\in\Omega_{0}^{r-1}(\mathcal{P})$, called the space of $(r;1)$-\textit{horizontal} forms. Its generalization, the space $\Omega^{r}_{s}(\mathcal{P})\subset\Omega^{r}(\mathcal{P})$ such that its elements $\theta$ satisfy $X\intprod\theta\in\Omega_{s-1}^{r-1}(\mathcal{P})$ for any $X\in\mathfrak{X}^{V}(\mathcal{P})$ is called the space of $(r;s)$-\textit{horizontal} forms.

As it is well known, the Lagrangian density $\mathcal{L}$, can be used to define a unique $n$-form associated to the first jet manifold $J^{1}E$, called the Cartan form \cite{saund,sardan2}, given by $\Theta_{L}=\frac{\partial\mathcal{L}}{\partial\phi^{a}_{\mu}}(d\phi^{a}-\phi_{\nu}^{a}dx^{\nu})\wedge V_{\mu}-\mathcal{L}V$, where $V_{\mu}:=\partial_{\mu}\intprod V$. By means of the De Donder-Weyl Legendre transformation (\ref{covleg}), the polymomentum phase-space $\mathcal{P}$ is also endowed with an $n$-form induced from $\Theta_{L}$ with the local representation
\begin{equation}\label{poincar}
\Theta_{DW}=\pi_{a}^{\mu}\,d\phi^{a}\wedge V_{\mu}-\Hdw V\,,
\end{equation}
known as the \textit{De Donder-Weyl Poincaré-Cartan form} \cite{katsup,bryant,forg,gold}. 
Even though the splitting of a differential form into its horizontal and vertical parts 
cannot  be done unless  the Ehresmann connection on the bundle is
specified,  we will explicitly consider this splitting
in the particular local representation~\eqref{poincar} as one may show  that the De Donder-Weyl field equations of a physical system may be encoded only in the vertical term of the De Donder-Weyl Poincaré-Cartan form \cite{got,katsup,kij,carin}. 
Indeed, despite the fact that the decomposition into vertical and horizontal parts may not be 
specified without a connection, the Poisson-Gerstenhaber
brackets associated to the field equations are connection independent as shown in~\cite{kan1,paufler}, and thus our 
choice of  a connection will not influence our treatment of the MacDowell-Mansouri
model below. 
In consequence, 
we will only consider the $(n;1)$-horizontal form $\Theta_{DW}^{V}\in\Omega_{1}^{n}(\mathcal{P})$ corresponding to the first term of (\ref{poincar}), as the second term corresponds to a purely horizontal form. 
The form $\Theta_{DW}^{V}$ may also be defined as an
equivalence class $\Theta_{DW}=\pi_{a}^{\mu}\,d\phi^{a}\wedge V_{\mu}$ modulo horizontal $n$-forms, endowing 
the polymomentum phase-space $\mathcal{P}$ with a canonical structure.
Now, given an arbitrary $p$-form in the polymomentum phase-space, namely, $\kan{F}{p}=\frac{1}{p!}F_{M_{1}\cdots M_{p}}dz^{M_{1}}\wedge\cdots\wedge dz^{M_{p}}$, where $z^{M}:=(x^{\mu},\phi^{a},\pi^{\mu}_{a})$ denotes the basis of $\mathcal{P}$, then we define its \textit{vertical derivative} by
\begin{equation}\label{vertder}
d^{V}\kan{F}{p}:=\frac{1}{p!}\partial_{v}F_{M_{1}\cdots M_{p}}dz^{v}\wedge dz^{M_{1}}\wedge\cdots\wedge dz^{M_{p}}\,,
\end{equation}
where $z^{v}:=(\phi^{a},\pi^{\mu}_{a})$ stands for the vertical coordinates in $\mathcal{P}$ \cite{kan1,kan10,kan11}. In this way, by taking the vertical derivative of $\Theta_{DW}^{V}$ we may define the $(n+1;2)$-horizontal form
\begin{equation}\label{mult}
\Omega_{DW}:=d^{V}\Theta_{DW}^{V}=d\pi_{a}^{\mu}\wedge d\phi^{a}\wedge V_{\mu}\,,
\end{equation}
called the \textit{polysymplectic form} \cite{kan2,gunter}. This $(n+1)$-form defines an analogue of the usual symplectic structure in the polymomentum phase-space. We aim to define the geometrical objects which, contracted with the polysymplectic form (\ref{mult}), maintain vertical information. With this in mind, let $\kan{X}{p}=\frac{1}{p!}X^{v\mu_{1}\cdots\mu_{p-1}}\partial_{v}\wedge\partial_{\mu_{1}}\wedge\cdots\wedge\partial_{\mu_{p-1}}$, with $0<p\leq n$, be a \textit{vertical multivector field} of degree $p$, \cite{forg,gold,helein,helein2}. We will call $\kan{X}{p}$ a \textit{Hamiltonian} multivector field, if there exists a unique $(n-p;0)$-horizontal form $\kan{F}{n-p}$ satisfying
\begin{equation}\label{Hamvec}
\kan{X}{p}\intprod\Omega_{DW}=d^{V}\kan{F}{n-p}\,.
\end{equation}
If this last condition is satisfied, then $\kan{F}{n-p}$ is called the \textit{Hamiltonian form} associated to the Hamiltonian multivector field $\kan{X}{p}$. This pairing between Hamiltonian forms and Hamiltonian
multivector fields induced from the polysymplectic form $\Omega_{DW}$, endows the polymomentum phase-space $\mathcal{P}$ with a Poisson-Gerstenhaber structure. Let $\kan{F}{p}\in\Omega_{0}^{p}(\mathcal{P})$, and $\kan{G}{q}\in\Omega_{0}^{q}(\mathcal{P})$ be Hamiltonian forms and $\kan{X}{n-p}_{\!\!F}$, and $\kan{X}{n-q}_{\!\!G}$ their respective Hamiltonian fields. The map $\dubbra\cdot,\cdot\dubarb :\Omega_{0}^{p}(\mathcal{P})\times\Omega_{0}^{q}(\mathcal{P})\longrightarrow\Omega_{0}^{p+q-n+1}(\mathcal{P})$ given by
\begin{equation}\label{GB}
\gerst{F}{G}{p}{q}:=(-1)^{n-p}\kan{X}{n-p}_{\!\!F}\intprod\kan{X}{n-q}_{\!\!G}\intprod\Omega_{DW}\,,
\end{equation}
is called the \textit{Poisson-Gerstenhaber bracket} \cite{kan1,kan10,kan11,kan2,helein,helein2}. Due to the degree corresponding to the contraction $\intprod$ of the right hand of (\ref{GB}), the Poisson-Gerstenhaber bracket is only defined for $p+q\geq n-1$. The commutation rule of this bracket is given by $\gerst{F}{G}{p}{q}=-(-1)^{|F||G|}\gerst{G}{F}{q}{p}$, where $|F|:=n-p-1$ and $|G|:=n-q-1$ are the degrees of the horizontal forms $\kan{F}{p}$ and $\kan{G}{q}$ with respect to the Poisson-Gerstenhaber bracket.  In consequence, this bracket structure results graded-commutative and, as a consequence, it satisfies the graded Jacobi identity $\jacger{F}{\gerst{G}{H}{q}{r}}{p}=\jacges{\gerst{F}{G}{p}{q}}{H}{r}+(-1)^{|F||G|}\jacger{G}{\gerst{F}{H}{p}{r}}{q}$. Finally, the Poisson-Gerstenhaber bracket satisfies a graded Leibniz rule 
\begin{equation}\label{leib}
\jacger{F}{\kan{G}{q}\bullet\kan{H}{r}}{p}=\gerst{F}{G}{p}{q}\bullet\,\,\kan{H}{r}+(-1)^{(n-q)|F|}\kan{G}{q}\,\,\bullet\gerst{F}{H}{p}{r}\,,
\end{equation}
where the map $\bullet :\Omega_{0}^{p}(\mathcal{P})\times\Omega_{0}^{q}(\mathcal{P})\longrightarrow\Omega_{0}^{p+q-n}(\mathcal{P})$, called the \textit{co-exterior product}, is given by
\begin{equation}\label{coex}
\kan{F}{p}\bullet\kan{G}{q}:=\ast^{-1}(\ast\kan{F}{p}\wedge\ast\kan{G}{q})\,,
\end{equation}
with $\ast$ being the Hodge dual defined over $M$. 
Note that within the construction used here there is no need 
to lift the Hodge star to the polymomentum phase-space $\mathcal{P}$, due to
the fact that both the Poisson-Gerstenhaber brackets and the 
co-exterior product are only defined on horizontal forms.
Equation \eqref{leib}, together with the commutation rule and the graded Jacobi identity make (\ref{GB}) a graded-Poisson bracket. 
This allows to define the fundamental Poisson-Gerstenhaber bracket relations among canonical 
conjugate variables by \cite{kan1,kan10,kan11}
\begin{equation}\label{canon}
\dubbra\pi^{\mu}_{a}V_{\mu},\phi^{b}\dubarb=\delta_{a}^{b}\,,\hspace{1cm}\dubbra\pi^{\mu}_{a}V_{\mu},\phi^{b}V_{\nu}\dubarb=\delta_{a}^{b}V_{\nu}\,,\hspace{1cm}\dubbra\pi^{\mu}_{a},\phi^{b}V_{\nu}\dubarb=\delta_{a}^{b}\delta_{\nu}^{\mu}\,.
\end{equation}
The co-exterior product induces a derivative operator called the \textit{total co-exterior differential} over horizontal forms \cite{kan1,kan10,kan11,kan2}. Let $\kan{F}{p}=\frac{1}{(n-p)!}F^{\mu_{1}\cdots\mu_{n-p}}(\partial_{\mu_{1}}\wedge\cdots\wedge\partial_{\mu_{n-p}})\intprod V$, be a horizontal $p$-form, then, its total co-exterior differential is defined as
\begin{equation}\label{coexdif}
\codif{F}{p}:=\frac{1}{(n-p)!}\partial_{v}F^{\mu_{1}\cdots\mu_{n-p}}\partial_{\mu}z^{v}\,dx^{\mu}\bullet(\partial_{\mu_{1}}\wedge\cdots\wedge\partial_{\mu_{n-p}})\intprod V+d^{h}\bullet\kan{F}{p}\,,
\end{equation}
where the last term is called the \textit{horizontal co-exterior differential} of $\kan{F}{p}$, namely,
\begin{equation}
d^{h}\bullet\kan{F}{p}=\frac{1}{(n-p)!}\partial_{\mu}F^{\mu_{1}\cdots\mu_{n-p}}dx^{\mu}\bullet(\partial_{\mu_{1}}\wedge\cdots\wedge\partial_{\mu_{n-p}})\intprod V\,.
\end{equation}
By means of equation (\ref{Hamvec}) and the definition of the Gerstenhaber bracket (\ref{GB}), the co-exterior differential of a Hamiltonian form $\kan{F}{p}$ can be written as
\begin{equation}\label{Hamgerst}
\codif{F}{p}=-\sigma(-1)^{n}\jacges{\Hdw}{F}{p}+d^{h}\bullet\kan{F}{p}\,,
\end{equation}
where $\sigma=\pm 1$ depends on the signature of the metric of the base space manifold $M$, $+1$ for Euclidean and $-1$ for Minkowski. Using the 
fundamental bracket relations (\ref{canon}), we obtain the field equations 
\begin{equation}\label{dweq}
\partial_{\mu}\phi^{a}
=
\dubbra\Hdw,\phi^a V_\mu\dubarb
=
\frac{\partial\Hdw}{\partial\pi^{\mu}_{a}}\,,\hspace{1cm}\partial_{\mu}\pi^{\mu}_{a}
=
\dubbra\Hdw,\pi^\mu_a V_\mu\dubarb
=
-\frac{\partial\Hdw}{\partial\phi^{a}}\,,
\end{equation}
known within this formalism as the \textit{De Donder-Weyl equations} \cite{sardan}. A direct calculation involving the definition of the De Donder-Weyl Hamiltonian can be used to show the equivalence between equations (\ref{dweq}) and the Euler-Lagrange field equations if the Lagrangian density satisfies the \textit{non-singular} condition $\mathrm{det}\left(\frac{\partial^{2}\mathcal{L}}{\partial\phi_{\mu}^{a}\partial\phi_{\nu}^{b}}\right)\neq 0$ \cite{got,sardan2}.   Even though the MacDowell-Mansouri model which we will analyse in the next section is described by a singular Lagrangian, the symmetries of the model allows 
to judiciously define the polymomenta
avoiding the standard treatment of
constraints, as described below.

\section{MacDowell-Mansouri gravity}\label{MM}

In this section we give a brief description of the system we will work with using the polysymplectic formalism. We will first introduce the MacDowell-Mansouri action and the relation it has with the Palatini action of gravity via a symmetry breaking of the gauge group $\mathrm{SO}(4,1)$. We will then treat the model using the polysymplectic formalism, closely focusing on the symmetry breaking process within the De Donder-Weyl Hamiltonian approach. 

\subsection{Lagrangian formalism}
The MacDowell-Mansouri action is based on a gauge theory with the gauge group $\mathrm{SO}(4,1)$ \cite{mm}, such that its correspondent Lie algebra admits the decomposition as vector spaces
\begin{equation}\label{split}
\mathfrak{so}(4,1)\simeq\mathfrak{so}(3,1)\oplus\mathbb{R}^{3,1}\,.
\end{equation}
Now, let  $M$ be the smooth base space manifold of the theory, and let the Lie algebra-valued $1$-form, $A\in\mathfrak{so}(4,1)\otimes\Omega^{1}(M)$, be the associated $\mathrm{SO}(4,1)$-connection, regarded as the gauge field. Decomposition (\ref{split}) splits the gauge field $A$ into an $\mathrm{SO}(3,1)$-connection $\omega$ and a \textit{coframe field} $e$, such that
\begin{equation}\label{conect}
A=\begin{pmatrix}
\omega & \frac{1}{l}e\\
-\frac{1}{l}e & 0
\end{pmatrix}\,,
\end{equation}
where $l$ is a constant chosen with units of length to be later related with the cosmological constant $\Lambda>0$. The coframe field $e$ is just the bundle morphism which makes the following diagram commutative, namely,
\begin{displaymath}
    \xymatrix{
        TM \ar[r]^e \ar[d]_\tau & \mathcal{T} \ar[dl]^{\pi} \\
        M        &  }
\end{displaymath}
where $\tau$ is the canonical tangent bundle projection, $\mathcal{T}$ is regarded as the internal space fibre bundle with local trivialization $\mathcal{T}=M\times\mathbb{R}^{3,1}$ and $\pi$ stands for its natural projection \cite{wise}. This coframe field induces a metric $g$ over the tangent bundle $TM$ by pulling back the metric $\eta$ of the internal space $\mathcal{T}$ such that $g=\eta(eu,ev)$ for any two vectors $u,v\in TM$.

The most relevant property of the gauge potential $A$, is that the associated gauge curvature $R=\mathrm{d}_{A}A:=\mathrm{d}A+A\wedge A$ also splits into an $\mathfrak{so}(3,1)$-valued $2$-form $F=\mathrm{d}_{\omega}\omega-\frac{1}{l^{2}}e\wedge e$ and the $\mathbb{R}^{3,1}$-valued $2$-form $\mathrm{d}_{\omega}e$, where symbolically $\mathrm{d}_{\omega}:=\mathrm{d}+\omega\wedge$ stands for the $\mathrm{SO}(3,1)$-gauge covariant exterior derivative,  such that
\begin{equation}\label{curv}
R=\begin{pmatrix}
F & \mathrm{d}_{\omega}e\\
-\mathrm{d}_{\omega}e & 0
\end{pmatrix}\,.
\end{equation}

In this way, the general MacDowell-Mansouri action with local gauge group $\mathrm{SO}(4,1)$, is given by
\begin{equation}\label{action}
\mathcal{S}[A]=\int_{\mathcal{C}}\mathrm{tr}\left(R\wedge\star R\right)\,,
\end{equation}
where the Hodge dual operator $\star$ and the trace 
are both to be taken with respect to the internal vector space $\mathfrak{so}(4,1)$.\footnote{This Hodge dual operator $\star$
must not be confused with $\ast$, appearing in
section~\ref{Mf},  which stands for the Hodge dual operator defined over $M$.} The MacDowell-Mansouri model of gravity is obtained by considering the projection of the curvature $R$ into the subalgebra $\mathfrak{so}(3,1)$, explicitly given by the action
\begin{equation}\label{mmac}
\mathcal{S}_{\mathsmaller{\mathsmaller{M\!M}}}[\omega,e]=\int_{\mathcal{C}}\mathrm{tr}\left(F\wedge\star F\right)\,,
\end{equation}
where, abusing notation, the Hodge dual and the trace are to be understood as acting within $\mathfrak{so}(3,1)$. It is relevant to notice that, by taking the projection of $R$, we have broken the $\mathrm{SO}(4,1)$ symmetry down to $\mathrm{SO}(3,1)$. The variation of the action (\ref{mmac}) with respect to the fields $\omega$ and $e$ results in the field equations
\begin{align}\label{fieldeq}
\mathrm{d}_{\omega}F&=0\,,\nonumber\\
e\wedge F&=0\,,
\end{align}
respectively. The first equation, using the definition of $F$, implies that $\mathrm{d}_{\omega}(e\wedge e)=0$, which is the torsion-free condition. The second expression in (\ref{fieldeq}) written in terms of the $\mathrm{SO}(3,1)$-curvature $\hat{R}:=\mathrm{d}_{\omega}\omega$, and fixing the constant $l$ in terms of the cosmological constant as $l^{2}=3/\Lambda$, yields the equation
\begin{equation}\label{eins}
e\wedge\hat{R}-\frac{\Lambda}{3}e\wedge e\wedge e=0\,,
\end{equation}
which is the coordinate free expression for Einstein's equations of gravity in terms of the connection $1$-form $\omega$ and the coframe $e$. 
The relation of the action $\mathcal{S}_{\mathsmaller{\mathsmaller{M\! M}}}$ with the classical  vierbein gravity formalism is obtained by means of the Palatini action 
\begin{equation}
\mathcal{S}_{\mathsmaller{Pal}}=\int_{\mathcal{C}}\mathrm{tr}\left(F\wedge\star F+\hat{R}\wedge\star\hat{R}\right)\,,
\end{equation}
in such a way that $\mathcal{S}_{\mathsmaller{Pal}}=\mathcal{S}_{\mathsmaller{\mathsmaller{M\!M}}}+T$, where $T$ stands for the topologically invariant term $\int \mathrm{tr}(\hat{R}\wedge\star\hat{R})$, whose variation classically  vanishes.

It is worth pointing out that a gravitational theory with cosmological constant term is obtained from (\ref{action}) only after breaking the $\mathrm{SO}(4,1)$-symmetry, as the variation of the original action gives the field equation $\mathrm{d}_{A}R=0$ which, upon the decomposition (\ref{curv}), results in the pair of equations
\begin{equation}\label{lagmans}
\mathrm{d}_{\omega}F+\frac{1}{l^{2}}e\wedge\mathrm{d}_{\omega}e=0\,,\hspace{1cm}\mathrm{d}_{\omega}^{2}e=0\,,
\end{equation}
for the $\mathfrak{so}(3,1)$ and $\mathbb{R}^{3,1}$ sectors, respectively. After breaking the $\mathrm{SO}(4,1)$ symmetry by projecting $\mathrm{d}_{A}R=0$ down to the $\mathfrak{so}(3,1)$ sector, the field equation obtained is the first expression in  \eqref{lagmans},  thus not recovering Einstein's equations, namely, $e\wedge F=0$. In other words, by breaking the $\mathrm{SO}(4,1)$-symmetry after the variation has been performed, the field equations of gravity cannot be recovered from the original action (\ref{lagmans}). This fact might be better understood by realizing that $\mathrm{d}_{\omega}F+\frac{1}{l^{2}}e\wedge\mathrm{d}_{\omega}e=0$ can be obtained by the variation of the inequivalent action 
\begin{equation}\label{altlag}
S[\omega]=\int_{\mathcal{C}}\mathrm{tr}\left(\hat{R}\wedge\star\hat{R}-\frac{1}{l^{2}}e\wedge e\wedge\star\hat{R}\right)\,,
\end{equation}
with respect to the field $\omega$, 
contrary to the action \eqref{mmac} which depends on both field variables $\omega$ and $e$.
As we will see below, the polysymplectic formalism behaves in a substantially different way under the symmetry breaking process due to the intrinsic definition of the polymomenta \eqref{polym}.

\subsection{De Donder-Weyl formulation}

In this section we will apply the polysymplectic formalism to the MacDowell-Mansouri action with $\mathrm{SO}(4,1)$-gauge symmetry. We will start by writing the action (\ref{action}) in local coordinates. Let $X_{IJ}\in\mathfrak{so}(4,1)$ be the generators of the correspondent $\mathfrak{so}(4,1)$ algebra, with $I=\{0,\dots 4\}$. In this way, the connection $A$ is given by $A=A_{\mu}^{IJ}dx^{\mu}\otimes X_{IJ}$, and the curvature $R$ takes the form $R=\frac{1}{2}R_{\mu\nu}^{IJ}\,dx^{\mu}\wedge dx^{\nu}\otimes X_{IJ}$ such that its components are given by definition, as
\begin{equation}
\label{curv2}
R_{\mu\nu}^{IJ}=2\partial_{[\mu}A^{IJ}_{\nu]}+A_{[\mu K}^{I}A_{\nu]}^{KJ}\,,
\end{equation}
with $\partial_{[\mu}A^{IJ}_{\nu]}=\frac{1}{2}(\partial_{\mu}A^{IJ}_{\nu}-\partial_{\nu}A^{IJ}_{\mu})$ denoting the antisymmetrization, and the internal contraction in the second term given by the $\mathfrak{so}(4,1)$-metric $\seta_{IJ}:=\mathrm{diag}(-1,1,1,1,1)_{IJ}$. Now, we will consider the action of the internal Hodge dual $\star$ as obtained through contraction with  $-Q_{IJKL}$, where these constants 
explicitly read
\begin{equation}\label{q}
Q_{IJKL}:=\frac{1}{2}(\seta_{IK}\seta_{JL}-\seta_{IL}\seta_{JK})\,.
\end{equation}
In this way, the action (\ref{action}) takes the local form
\begin{equation}\label{locac}
\mathcal{S}[A]=-\frac{1}{4}\int_{\mathcal{C}}R^{IJ}_{\mu\nu}R^{KL}_{\rho\sigma}\epsilon^{\mu\nu\rho\sigma}Q_{IJKL}\,V\,,
\end{equation}
where the $\epsilon^{\mu\nu\rho\sigma}$ is the usual Levi-Civita alternating symbol for the base space manifold $M$. Following (\ref{polym}), the associated polymomenta are given by
\begin{equation}\label{Poly}
\Pi_{IJ}^{\mu\nu}=-\epsilon^{\mu\nu\rho\sigma}Q_{IJKL}R^{KL}_{\rho\sigma}\,.
\end{equation}
Noticing that $Q^{IJMN}Q_{IJKL}=\delta_{[K}^{M}\delta_{L]}^{N}$, and using (\ref{curv2}), we can write the antisymmetric derivatives of $A$ in terms of the polymomenta, thus we have that $\partial_{[\mu}A_{\nu]}^{IJ}=-\frac{1}{8}\epsilon_{\mu\nu\rho\sigma}Q^{IJKL}\Pi_{KL}^{\rho\sigma}-\frac{1}{2}A_{[\mu M}^{I}A_{\nu]}^{MJ}$.  From the definition of the polymomenta, we also may see that the symmetric part vanishes, thus implying
the presence of constraints $\Pi^{(\mu\nu)}_{IJ}\approx 0$.   A straightforward calculation shows that the polymomenta 
associated to the symmetric derivatives of the connection is  also divergenceless and, 
as demonstrated in~\cite{eslava}, within the polysymplectic formalism one may  redefine the
polymomenta in order to circumvent the presence of 
any symmetric part, thus avoiding the standard 
treatment of constraints in the symmetric sector of the configuration space.

Then, the De Donder-Weyl Hamiltonian  (\ref{covleg}), takes the local form
\begin{equation}\label{dwham}
\Hdw=-\frac{1}{16}\epsilon_{\mu\nu\rho\sigma}Q^{IJKL}\Pi_{IJ}^{\mu\nu}\Pi_{KL}^{\rho\sigma}-\frac{1}{2}\Pi_{IJ}^{\mu\nu}A_{[\mu K}^{I}A_{\nu]}^{KJ}\,,
\end{equation}
and the De Donder-Weyl equations (\ref{dweq}) for this Hamiltonian read
\begin{align}\label{eq1}
\partial_{\mu}A_{\nu}^{IJ}&=\dubbra\Hdw,A_{\nu}^{IJ}V_{\mu}\dubarb=-\frac{1}{8}\epsilon_{\mu\nu\rho\sigma}Q^{IJKL}\Pi_{KL}^{\rho\sigma}-\frac{1}{2}A_{[\mu M}^{I}A_{\nu]}^{MJ}\,,\nonumber\\
\partial_{\mu}\Pi_{IJ}^{\mu\nu}&=\dubbra\Hdw,\Pi^{\mu\nu}_{IJ}V_{\mu}\dubarb=-A_{\mu[I}^{K}\Pi_{J]K}^{\mu\nu}\,,
\end{align}
where the first expression is equivalent to an identity from the definition of the polymomenta (\ref{Poly}). Notice that both equations reduce to $\mathrm{d}_{A}R=0$ when substituted one into the other, demonstrating the equivalence of the De Donder-Weyl and Lagrangian field equations.

Now, we proceed with the decomposition (\ref{split}) without breaking the $\mathrm{SO}(4,1)$-symmetry, that is, considering all the components of $A$ and $\Pi$. Let us consider the set of indices for the internal subalgebra $\mathfrak{so}(3,1)$, as the lower case Latin indices $a:=\{0,\dots, 3\}$, such that $I=\{a,4\}$. From relation (\ref{conect}), we may define the local components of the connection $A$ by
\begin{equation}\label{conn31}
\omega_{\mu\,b}^{a}:=A_{\mu\,b}^{a}\,,\hspace{1cm}\frac{1}{l}e_{\mu}^{a}:=A_{\mu\,4}^{a}\,,
\end{equation}
where the antisymmetry of the connection $A$ implies that $\frac{1}{l}e_{\mu\,a}=-A_{\mu\,a}^{4}$. In a similar fashion, the decomposition of the internal space induces a splitting of the components of the polymomenta, thus let us define
\begin{equation}\label{Poly31}
\pi_{ab}^{\mu\nu}:=\Pi_{ab}^{\mu\nu}\,,\hspace{1cm}\frac{1}{l}p_{a}^{\mu\nu}:=\Pi_{a\,4}^{\mu\nu}\,.
\end{equation}
Also, the decomposition of the constants $Q_{IJKL}$, from definition (\ref{q}), satisfy $Q_{abc4}=0$ and $Q_{a4b4}=\frac{1}{2}\mathlarger{\eta}_{ab}$, while the
rest vanish. In this way, we can explicitly write the decomposition of the De Donder-Weyl equations by fixing indices in both (\ref{eq1}), such that the $\mathfrak{so}(3,1)$ part is obtained with $I=a$ and $J=b$, and the $\mathbb{R}^{3,1}$ part with $I=a$ and $J=4$. The former index fixing, using the definitions (\ref{conn31}) and (\ref{Poly31}), yields the equations
\begin{align}\label{dweqso31}
\pi_{ab}^{\mu\nu}&=-\epsilon^{\mu\nu\rho\sigma}Q_{abcd}\left(2\partial_{[\rho}\omega_{\sigma]}^{cd}+\omega_{[\rho\,i}^{c}\omega_{\sigma]}^{id}-\frac{1}{l^{2}}e_{[\rho}^{c}e_{\sigma]}^{d}\right)\,,\nonumber\\
\partial_{\mu}\pi_{ab}^{\mu\nu}&=-\omega_{\mu[a}^{c}\pi_{b]c}^{\mu\nu}+\frac{1}{l^{2}}e_{\mu[a}p_{b]}^{\mu\nu}\,,
\end{align}
and in a similar way, the $\mathbb{R}^{3,1}$ part reads 
\begin{align}\label{dweqr31}
p_{a}^{\mu\nu}&=-\epsilon^{\mu\nu\rho\sigma}\mathlarger{\eta}_{ab}\left(2\partial_{[\rho}e_{\sigma]}^{b}+\omega_{[\rho\,c}^{b}e_{\sigma]}^{c}\right)\,,\nonumber\\
2\partial_{\mu}p^{\mu\nu}_{a}&=\omega_{\mu\,a}^{b}p_{b}^{\mu\nu}-e_{\mu}^{c}\pi_{ac}^{\mu\nu}\,.
\end{align}

Substituting the first line in (\ref{dweqso31}) into the second line, and by using the definition of the projection $F$ \eqref{curv} of the curvature $R$ down to $\mathfrak{so}(3,1)$, we obtain the relation
\begin{equation}\label{dweq1}
\mathrm{d}_{\omega}F^{ab}=\frac{1}{l^{2}}e^{a}\wedge(\mathrm{d}_{\omega}e)^{b}\,,
\end{equation}
and in a similar way, by substituting the first line in (\ref{dweqr31}) into the second line, and by using \eqref{curv}, we arrive at the expression
\begin{equation}\label{dweq2}
\mathrm{d}_{\omega}(\mathrm{d}_{\omega}e)^{a}=-e^{b}\wedge F_{\,\,\,\,\,b}^{a}\,.
\end{equation}
Equations~(\ref{dweq1}) and~(\ref{dweq2}) may be thought of as a consequence of the structural Cartan equations~\cite{cartan,cartan2}.  
It is relevant to notice that this last equation differs from the second Lagrangian field equation in (\ref{fieldeq}) in an important way. To see this clearly, let us note that the $\mathrm{SO}(4,1)$-symmetry breaking in this formalism is achieved by requiring that $p_{a}^{\mu\nu}=0$, which is equivalent to taking only the projection $F$ of $R$ in the Lagrangian approach. With this condition, the De Donder-Weyl equations, (\ref{dweq1}) and (\ref{dweq2}), read $\mathrm{d}_{\omega}F^{ab}=0$ and $e^{b}\wedge F_{\,\,\,\,\,b}^{a}=0$, respectively, which turn out to be exactly the field equations of the $\mathrm{SO}(3,1)$ MacDowell-Mansouri action, (\ref{mmac}). This means that, in contrast to the Lagrangian approach, the De Donder-Weyl formalism appears to allow the symmetry breaking after the variation of $\Hdw$ has been done and still reproduce the same physical behaviour.  At first glance this could sound contradictory due to the fact that the De Donder-Weyl equations are equivalent to the Euler-Lagrange field equations, as in this particular model, for the reasons we have explained, this seems  not to be the case. There is, however, a simple but quite interesting aspect of the De Donder-Weyl formulation which explains this discrepancy with the Lagrangian approach. The polymomenta \eqref{polym} can be defined as the variation of the action \eqref{action} with respect to the derivative coordinates $\phi_{\mu}^{a}$ or, for this particular model, with respect to $\partial_{\mu}\omega_{\nu}^{ab}$ and $\partial_{\mu}e_{\nu}^{a}$. Thus, by breaking the $\mathrm{SO}(4,1)$ symmetry in \eqref{dweqso31} and \eqref{dweqr31}, the intrinsic variation with respect to both variables present in the polymomenta remains after taking the projection down to $\mathfrak{so}(3,1)$, while in the Lagrangian case, when we broke the symmetry in the equation $\mathrm{d}_{A}R=0$, the variation with respect to the field $\partial_{\mu}e_{\nu}^{a}$ is lost in the process, thus, giving different equations, namely, \eqref{lagmans}, corresponding to the variation of the inequivalent action \eqref{altlag}.
 For these reasons, in the polysymplectic formalism, the De Donder-Weyl equations result invariant if the symmetry breaking process is performed either before or after the variation of the De Donder-Weyl Hamiltonian. Indeed, by considering the De-Donder Weyl Hamiltonian (\ref{dwham}), decomposed as
\begin{eqnarray}\label{dwhamdec}
\Hdw=&-&\frac{1}{16}\epsilon_{\mu\nu\rho\sigma}\epsilon^{ijab}\epsilon_{ij}^{\,\,\,\,\,cd}\pi_{ab}^{\mu\nu}\pi_{cd}^{\rho\sigma}-\frac{1}{2}\pi_{ab}^{\mu\nu}\left(\omega^{a}_{[\mu\,c}\omega^{cb}_{\nu]}-\frac{1}{l^{2}}e^{a}_{[\mu}e^{b}_{\nu]}\right)\nonumber\\
&-&\frac{1}{8l^{2}}\epsilon_{\mu\nu\rho\sigma}\mathlarger{\eta}^{ab}p_{a}^{\mu\nu}p_{b}^{\rho\sigma}-\frac{1}{l^{2}}p_{a}^{\mu\nu}\omega^{a}_{[\mu\,b}e^{b}_{\nu]}\,,
\end{eqnarray}
where we have used the fact that $Q^{abcd}=\epsilon^{ijab}\epsilon_{ij}^{\,\,\,\,\,cd}$, we now can first apply the symmetry breaking and then variate the $\mathrm{SO}(3,1)$ Hamiltonian. In order to obtain the field equations from the decomposed Hamiltonian \eqref{dwhamdec}, we first need to determine which fields stand as canonical variables. To do so, we use the fundamental Poisson-Gerstenhaber bracket relations (\ref{canon}) for the $\mathrm{SO}(4,1)$ Hamiltonian (\ref{dwham}), namely,  $e\mathrm{e}$
\begin{align}
\label{commut}
\dubbra\Pi_{IJ}^{\mu\nu}V_{\nu},A^{KL}_{\rho}\dubarb
& =  \delta^{\mu}_{\rho}\delta_{I}^{[K}\delta_{J}^{L]}\,, 
\nn\\
\dubbra\Pi_{IJ}^{\mu\nu}V_{\nu},A^{KL}_{\rho}V_{\sigma}\dubarb
& =  
\delta^{\mu}_{\rho}\delta_{I}^{[K}\delta_{J}^{L]}V_{\sigma}\,,
\nn\\
\dubbra\Pi_{IJ}^{\mu\nu},A^{KL}_{\rho}V_{\sigma}\dubarb
& =  
\delta^{\mu}_{\rho}\delta^{\nu}_{\sigma}\delta_{I}^{[K}\delta_{J}^{L]}\,,
\end{align}
and again, by fixing indices for the $\mathfrak{so}(3,1)$ and the $\mathbb{R}^{3,1}$ parts, and using the variables defined in (\ref{conn31}) and (\ref{Poly31}), we obtain the fundamental bracket relations
\begin{equation}\label{deccommut}
\begin{array}{ll}
\dubbra\pi_{ab}^{\mu\nu}V_{\nu},\omega_{\rho}^{cd}\dubarb=\delta_{\rho}^{\mu}\delta_{a}^{c}\delta_{b}^{d}\,,&\dubbra p_{a}^{\mu\nu}V_{\nu},e_{\rho}^{b}\dubarb=\delta_{\rho}^{\mu}\delta_{a}^{b}\,,\\
\dubbra\pi_{ab}^{\mu\nu}V_{\nu},\omega_{\rho}^{cd}V_{\sigma}\dubarb=\delta_{\rho}^{\mu}\delta_{a}^{c}\delta_{b}^{d}V_{\sigma}\,,&\dubbra p_{a}^{\mu\nu}V_{\nu},e_{\rho}^{b}V_{\sigma}\dubarb=\delta_{\rho}^{\mu}\delta_{a}^{b}V_{\sigma}\,,\\
\dubbra\pi_{ab}^{\mu\nu},\omega_{\rho}^{cd}V_{\sigma}\dubarb=\delta_{\rho}^{\mu}\delta_{\sigma}^{\nu}\delta_{a}^{c}\delta_{b}^{d}\,,&\dubbra p_{a}^{\mu\nu},e_{\rho}^{b}V_{\sigma}\dubarb=\delta_{\rho}^{\mu}\delta_{\sigma}^{\nu}\delta_{a}^{b}\,,
\end{array}
\end{equation}
which show that the pairs $(\omega,\pi)$ and $(e,p)$ assume the role of canonical variables with respect to the Poisson-Gerstenhaber bracket after the decomposition of the internal algebra. Now, if we break the symmetry at this point by imposing once again $p_{a}^{\mu}=0$, then the $\mathrm{SO}(3,1)$ De Donder-Weyl Hamiltonian consist only of the first line of (\ref{dwhamdec}), and thus, the De Donder-Weyl equations for this case are given by
\begin{align}
\partial_{\mu}\omega_{\nu}^{ab}=\dubbra\Hdw,\omega_{\nu}^{ab}\,V_{\mu}\dubarb&=-\frac{1}{8}\epsilon_{\mu\nu\rho\sigma}Q^{abcd}\pi_{cd}^{\rho\sigma}-\frac{1}{2}\left(\omega_{[\mu\,c}^{a}\omega_{\nu]}^{cb}-\frac{1}{l^{2}}e_{[\mu}^{a}e_{\nu]}^{b}\right)\,,\\
\partial_{\mu}\pi^{\mu\nu}_{ab}=\dubbra\Hdw,\pi^{\mu\nu}_{ab}\,V_{\mu}\dubarb&=-\omega_{\mu[a}^{c}\pi_{b]c}^{\mu\nu}\,,\\
0=\dubbra\Hdw, p_{\nu}^{a}\,V_{\mu}\dubarb&=-e_{\mu}^{c}\pi_{ac}^{\mu\nu}\,,
\end{align}
which, after explicitly writing them in terms of $F$, reduce to (\ref{fieldeq}), giving the same physical behaviour as before. In contrast to the Lagrangian approach, the De Donder-Weyl formalism allows to break the $\mathrm{SO}(4,1)$ symmetry either before or after determining the field equations. As we have mentioned above, thanks to the introduction of polymomenta the polysymplectic formalism allows to explicitly maintain
the invariance of the De Donder-Weyl Hamiltonian under the symmetry breaking.

\section{Conclusions}
\label{sec:conclu}
One of the main reasons why the De Donder-Weyl formulation gives an interesting approach to physical
models is its intrinsic spacetime covariant structure which easily fits for the systems with strong geometric content. Although the foundations for this formalism were introduced almost a century ago,
the introduction of the relevant polysymplectic structure, the  construction of the Poisson-Gerstenhaber
defined on forms, and applications of this formalism to  compelling physical systems are relatively new~\cite{kan1,kan10,kan11,kan2,kan6}, as only a few concrete examples using this method have been developed~\cite{kan1,kan10,kan11,kan2,helein,helein2,helein3,eslava,vey,kan7,kan8,kan9}. 
Some other recent references where
other alternative geometric formalisms are addressed for models in General Relativity may be found in~\cite[and references therein]{capri,gaset}.
In this sense, 
one of our major motivations for this paper was to extend the repertoire of physically relevant models studied under this De Donder-Weyl Hamiltonian approach.
In particular, we analysed the well-known MacDowell-Mansouri model of gravity~\cite{mm}, for which we developed 
both the Lagrangian and polysymplectic formulations.
At both levels, we found the correspondent forms of the field equations.   As we have shown throughout our analysis, 
the inherently spacetime covariant nature of the polymomenta resulted completely relevant in order to 
study the symmetry breaking process within the polysymplectic formalism.
Indeed, at the Lagrangian level the variational process 
does not commute with the symmetry breaking 
$\mathrm{SO}(4,1)\rightarrow\mathrm{SO}(3,1)$, 
yielding two inequivalent sets of field equations, namely \eqref{fieldeq} and \eqref{lagmans}, which 
were obtained by performing these two procedures in 
different order.
However, within the polysymplectic approach we noticed that the symmetry breaking leaves invariant the emerging De Donder-Weyl equations that follow from  
the Poisson-Gerstenhaber bracket. 
To understand this, we noticed that the symmetry 
breaking at the polysymplectic level includes variations with respect to all the polymomenta, and 
these polymomenta precisely include spacetime derivatives of the fields in each of the 
sectors in which the gauge algebra $\mathfrak{so}(4,1)$
is decomposed.  As discussed above, this is not the case
at the Lagrangian level.

The fact that the De Donder-Weyl Hamiltonian preserves the information of the variation of the action in 
a completely spacetime covariant manner encourages the analysis of some other physical models within the polysymplectic approach. 
A natural question in this direction is to address  whether other similar models with broken symmetry may behave in an analogous way to the MacDowell-Mansouri 
gravity model, so that the analysis presented in the paper could also serve as
a starting point of the study of other physical models with a symmetry breaking.

\section*{Acknowledgements}

The authors would 
like to thank Eslava del R\'io for collaboration and discussions. AM
acknowledges financial support from 
CONACYT-Mexico under project 
CB-2014-243433.

\end{document}